\documentclass[preprint,pre,onecolumn]{revtex4}

\usepackage{amsfonts}
\usepackage{amsmath}
\usepackage{amssymb}
\usepackage{graphicx}

\begin{document}

\preprint{}
\title{Self-assembling DNA-caged particles: nanoblocks for hierarchical
self-assembly}
\author{Nicholas A. Licata$^{1,2}$ and Alexei V. Tkachenko$^{1}$}
\affiliation{$^{1}$Department of Physics and Michigan Center for Theoretical Physics,
University of Michigan, 450 Church Street, Ann Arbor, Michigan 48109}
\affiliation{$^{2}$Max Planck Institue for the Physics of Complex Systems, N\"{o}thnitzerstrasse 38, 01187 Dresden, Germany}

\begin{abstract}
DNA is an ideal candidate to organize matter on the nanoscale, primarily due
to the specificity and complexity of DNA based interactions. \ Recent
advances in this direction include the self-assembly of colloidal crystals
using DNA grafted particles. \ In this article we theoretically study the
self-assembly of DNA-caged particles. \ These nanoblocks combine DNA grafted
particles with more complicated purely DNA\ based constructs. \
Geometrically the nanoblock is a sphere (DNA grafted particle) inscribed
inside a polyhedron (DNA cage). \ The faces of the DNA cage are open, and
the edges are made from double stranded DNA. \ The cage vertices are
modified DNA junctions. \ We calculate the equilibriuim yield of
self-assembled, tetrahedrally caged particles, and discuss their stability
with respect to alternative structures. \ The experimental feasability of
the method is discussed. \ To conclude we indicate the usefulness of
DNA-caged particles as nanoblocks in a hierarchical self-assembly strategy.
\ 
\end{abstract}

\maketitle

\section{Introduction}

DNA is one of the most celebrated tools in the nanoscience toolbox. \ This
approach was pioneered in the laboratory of N. Seeman, where some of the
first schemes for building nanostructures from specially engineered
oligonucleotide sequences were proposed. \ A number of objects have been
successfully constructed, including DNA\ cubes, multiple armed DNA
junctions, DNA crystals, and DNA\ knots (\cite{natreview},\cite{nucleic},%
\cite{nanotech},\cite{bottomup}).\ There have been several 
recent experimental advances in this direction (\cite{Bipyramid},\cite{Chiral}), 
including the encapsulation of a single molecule inside 
a DNA tetrahedron \cite{Encapsulate}. \ There has also been a surge of
interest in utilizing the specific interactions of complementary
single-stranded DNA (ssDNA) to organize particles on the nanoscale. \ One
recent advance in this direction is the self-assembly of three dimensional
body centered cubic crystals from DNA grafted nanoparticles (\cite%
{colloidalcrystal},\cite{mirkincrystal}). \ Up until this point, most of the
studies reported the formation of small clusters or random aggregation of
particles, as opposed to the self-assembly of ordered structures (\cite%
{micelle},\cite{chaikin},\cite{crocker}). \ 

\begin{figure}[tbp]
\includegraphics[width=5.0548in,height=3.8009in]{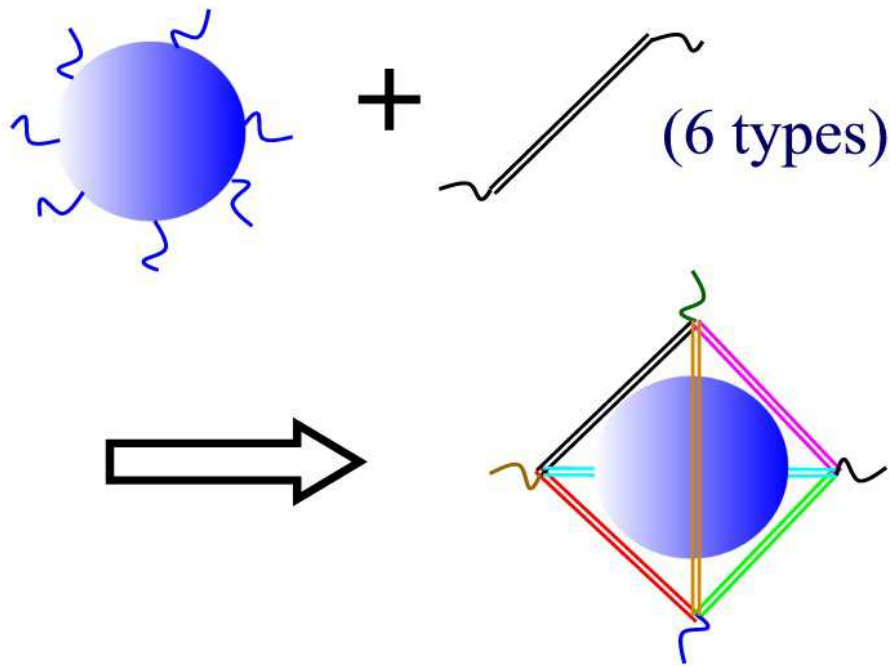}
\caption{(Color online). Graphical
depiction of the scheme for self-assembling DNA-caged particles. The cage edges
are constructed from dsDNA rods with terminal ssDNA sequences on either end.}
\label{cage}
\end{figure}

The potential complexity of DNA based interactions provides a means to
design significantly more complicated nanoblocks. \ In this paper we propose
a method to self-assemble DNA-caged particles (see Fig. \ref{cage}). \ These
nanoblocks are composite materials which are constructed by combining DNA
grafted nanoparticles with specially designed DNA sequences. \ Geometrically
the nanoblock is a sphere (DNA grafted nanoparticle) inscribed inside of a
polyhedron (DNA cage). \ The polyhedron faces are open, and the cage edges
are made of double-stranded DNA (dsDNA). \ The cage vertices are modified
DNA junctions (\cite{junctionflex}, \cite{designjunction}). \ Each vertex of
the cage carries a unique ssDNA\ sequence available for hybridization. \
This vertex ``coloring'' makes these nanoblocks ideal candidates as building
blocks for hierarchical self-assembly strategies. \ 

The plan for the paper is the following. \ We first introduce of our
self-assembly proposal. \ Details of the proposal are discussed for a
particular implementation in which the DNA cage is a regular tetrahedron. \ 
We theoretically calculate the melting profile for the DNA\ cage
self-assembled around the DNA\ grafted particle. \ We demonstrate an
equilibrium regime in which the DNA-caged particle is the dominant structure
in solution, and discuss its stability with respect to alternative
structures. \ We conclude by discussing how DNA-caged particles could be
used as the building blocks in a hierarchical self-assembly strategy. \ 

\section{Self-assembly proposal}

In this section we discuss the details of our self-assembly proposal. \ The
proposal is to self-assemble DNA-caged particles by combining DNA grafted
nanoparticles with rod-like DNA\ linkers. \ There is only one type of
particle, i.e. all of the ssDNA grafted onto the nanoparticle surface have
the same nucleotide sequence. \ The rod-like DNA linkers are dsDNA, but each
end of the rod terminates in a ssDNA\ sequence. \  The dsDNA rods can bend significantly 
when their length $L$ is comporable to the persistence length $l_{p} \simeq$ 50nm for dsDNA \cite{Flexibility}. 
  \  In what follows we consider the case $L \ll l_{p}$ and 
treat the dsDNA as rigid rods. \  There are $n$ types of rod-like DNA
linkers, since the terminal ssDNA sequences on each linker are unique. \ The
particle and linkers are all of the components necessary for the
self-assembly proposal. \ The number of types of linkers is determined by
the cage architecture, in general there will be one type for each edge of
the cage. \ We now turn to discuss how the cage can be assembled from the
DNA\ linkers. \

For the sake of concreteness we will consider a particular implementation of
this idea in which the DNA cage is a regular tetrahedron. \ In this case
there are $n=6$ types of DNA rods, one for each edge of the tetrahedron. \ 

\begin{figure}[tbp]
\includegraphics[width=5.0548in,height=3.8009in]{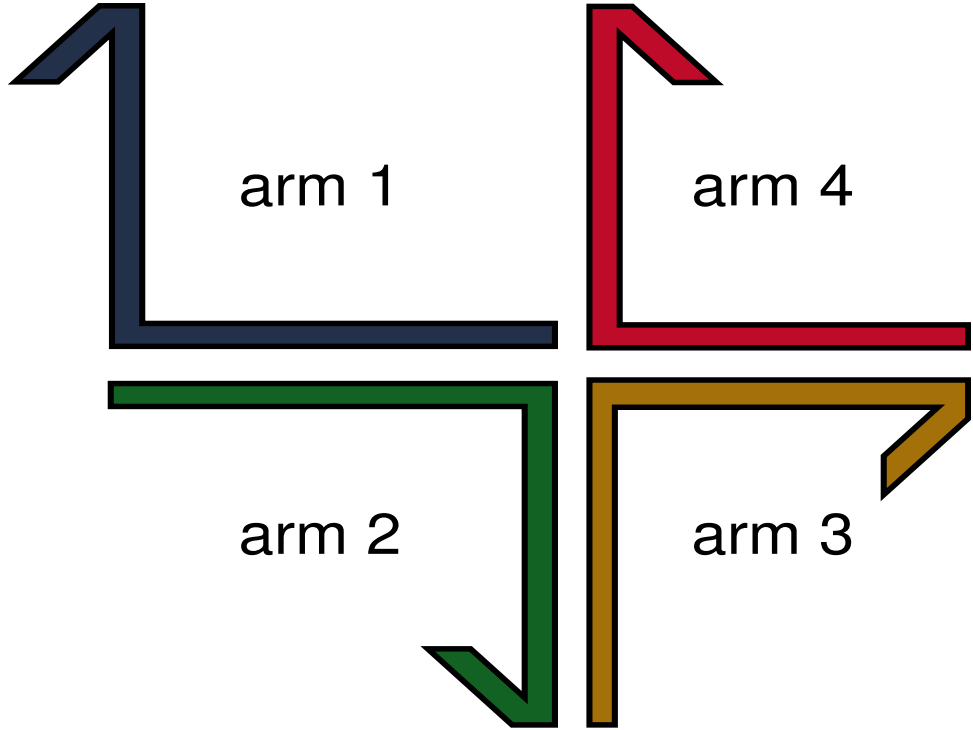}
\caption{(Color online). The vertex architecture is a modified $4$ 
arm DNA junction. \ The arrowheads label the $3^{\prime }$ end of the
ssDNA arms. \ Note that the portion of arm $1$ which is normally
complementary to arm $4$ is missing. \ Hence the $3^{\prime }$ end of arm $4$
provides the vertex with a unique ``color", i.e. a ssDNA sequence available
for hybridization. \  }
\label{dnajunction}
\end{figure}

These rods can be joined to assemble the cage in the following manner.
\ To construct each vertex of the tetrahedron, four ssDNA sequences must be joined. \ 
Three of these ssDNA sequences are the terminal ssDNA sequences of the rod-like DNA linkers.  The fourth ssDNA sequence
comes from the ssDNA grafted onto the particle surface, which binds the particle to the cage. \ 
\ The DNA architecture that accomplishes this task is
known as a four arm DNA\ junction (see Fig. \ref{dnajunction}). \ These junctions have
been studied extensively, and the sequences can be designed so that the
vertex is stable \cite{junctionflex,designjunction,GIDEON}. \

\begin{figure}[tbp]
\includegraphics[width=5.0548in,height=3.8009in]{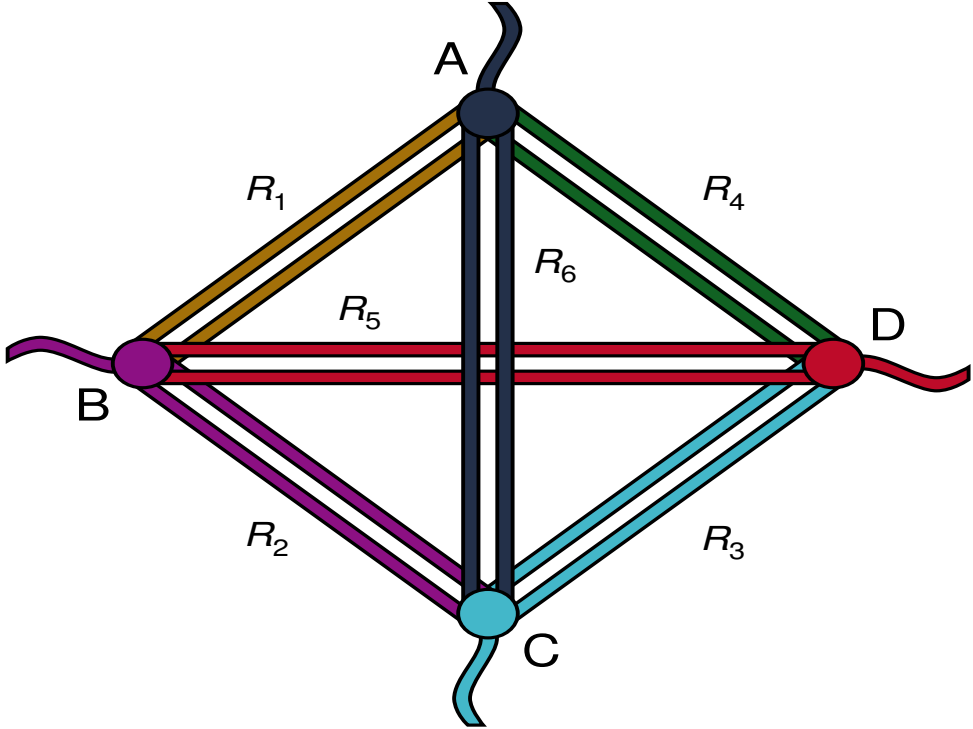}
\caption{(Color online).  The proposed
sequence assignment to the ssDNA ends of the DNA rods which
self-assembles into the tetrahedral cage.  \ If $S$ denotes a dsDNA spacer,
the sequence assignments for the $6$ rods are: $R_{1}=A_{2}-S-B_{2}$, $%
R_{2}=B_{4}-S-C_{4}$, $R_{3}=C_{2}-S-D_{2}$, $R_{4}=A_{4}-S-D_{4}$, $%
R_{5}=B_{3}-S-D_{1}$, $R_{6}=A_{3}-S-C_{3}$. \ The detailed structure of the vertex architecture 
(the circles in this diagram) is presented in Fig. \ref{dnajunction}. \  }
\label{cagecoloring}
\end{figure}

The problem is now to assign particular sequences to the terminal ssDNA sequences of the DNA rods which result
in the desired tetrahedral cage, taking into account the proposed vertex
architecture. \ One such assignment is proposed below in Fig. \ref{cagecoloring}. \ 
We now provide an explanation of how Figs. \ref{dnajunction} and \ref{cagecoloring} can 
be read together to understand the cage construction. \ 

Examine vertex $A$ in Fig. \ref{cagecoloring}. 
\ We can see that rods $R_{1}$, $R_{4}$, and $R_{6}$ are joined together at this vertex. \ Let 
$A_{n}$ denote the nucleotide sequence which plays the role of arm $n$ (see Fig. \ref{dnajunction}) in vertex $A$, where 
$n \in \{1,2,3,4\}$. \ 
The sequence assignments in the caption of Fig. \ref{cagecoloring} tell which rod provides each arm of the DNA junction. \ 
For example, $R_{1}=A_{2}-S-B_{2}$ means that rod $R_{1}$ provides arm $2$ of vertex $A$ and arm $2$ of vertex $B$. \ 
For vertex $A$, rod $R_{1}$ provides arm $2$, $R_{4}$ provides arm $4$, and $R_{6}$ provides arm $3$. \ Only arm $1$ remains, 
which is provided by the ssDNA grafted onto the particle. \ In addition, since all of the ssDNA grafted onto
the particle have the same sequence, by performing this enumeration procedure for each vertex we can see that the following four 
sequences are identical: $A_{1}=B_{1}=C_{1}=D_{3}$. \ 

Since it may be difficult (e.g. for steric
reasons) to introduce the particle into the fully assembled cage, we would
like for the particle to assist in the cage building process. \ This has
been explicitly taken into account in the sequence designation process. \
Note that rod $R_{4}$ cannot bind at vertex $D$ in the absence of the
particle, since it hybridizes to arm $3$ of the vertex (which comes from the
particle). \

With the basic
framework in hand, the next task is to determine the relative abundance of
the various structures that form in a solution of DNA linkers and
DNA-grafted nanoparticles. \ A similar type of analysis has been performed
in our related work on DNA-grafted nanoparticles (\cite{nanoclusters},\cite%
{statmech}). \ In the next section we calculate the equilibrium yield for a
variety of these structures. \ If the self-assembly process is
experimentally feasible we should be able to demonstrate a regime in which
our nanoblock, a single particle surrounded by a fully assembled DNA cage,
is the dominant structure in solution. \ 

\section{DNA cage melting}

\begin{figure}[tbp]
\includegraphics[width=5.0548in,height=3.8009in]{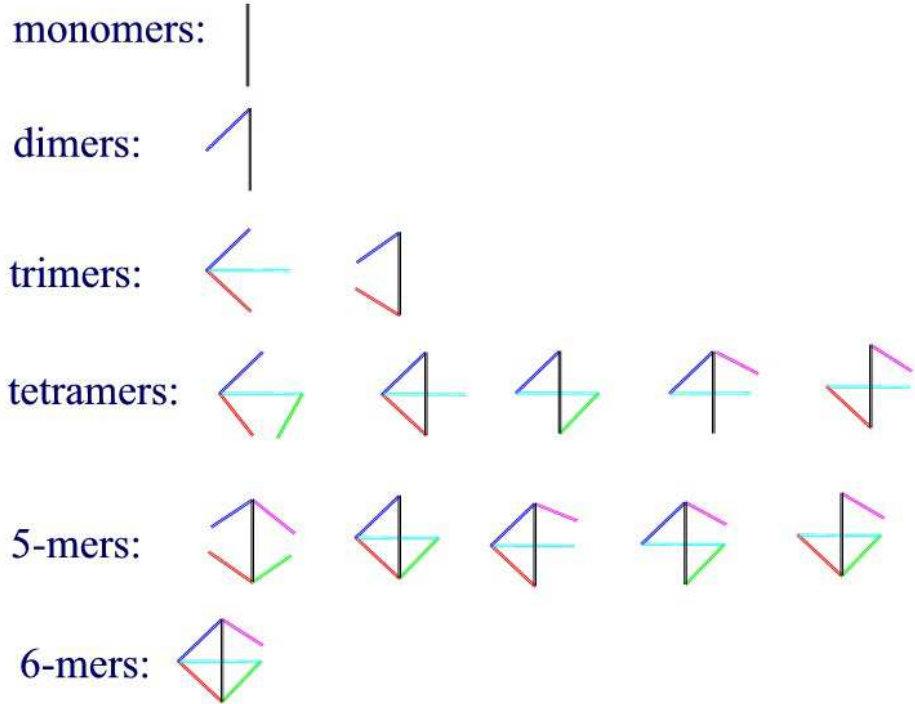}
\caption{(Color online).  The subsets of the DNA cage (tetrahedron) which can be formed in
the absence of the particle. \ For each topologically distinct diagram only
one variety is shown. \  }
\label{mers}
\end{figure}

We first determine the melting profile for DNA cages in the absence of the
nanoparticle. \ By taking the proposed vertex numbering scheme (see Fig. \ref%
{cagecoloring}) into account, we can enumerate all of the possible DNA\
structures which can form in solution (see Fig. \ref{mers}). \ 

Let $c_{i}$ denote the concentration of linker $R_{i}$. \ In what
follows $c_{o}=1M$ is the standard reference concentration. \ In addition,
we use natural units where the Boltzmann constant $k_{B}=1$. \ In
equilibrium the chemical potential of the various phases will be the same. \
The chemical potential has a contribution from the entropy of dilution, and
the effective hybridization free energy for creating the the DNA connections
at the vertex. \ For example, consider the reaction in which DNA linker rods 
$R_{1}$ and $R_{6}$ hybridize to form a dimer. \ Equilibrating the chemical
potentials yields the following equation:%
\begin{equation}
T\log \left( \frac{c_{1}}{c_{o}}\right) +T\log \left( \frac{c_{6}}{c_{o}}%
\right) =T\log \left( \frac{c_{16}}{c_{o}}\right) +\varepsilon _{16}
\end{equation}%
Here $\varepsilon _{16}$ is the free energy for the formation of the dimer
pair from the two monomers. \ In this case $\varepsilon _{16}$ is simply the
hybridization free energy associated with joining rods $R_{1}$ and $R_{6}$
together at vertex $A$. \ The resulting concentration of the dimer $c_{16}$
is then 
\begin{equation}
c_{16}=\frac{c_{1}c_{6}}{c_{o}}\exp \left( \frac{-\varepsilon _{16}}{T}%
\right)
\end{equation}%
The total concentration of dimers $\overline{C}_{2}$ is determined by
considering all of the possible dimer varieties. \ 
\begin{equation}
\overline{C}_{2}=c_{16}+c_{46}+c_{15}+c_{25}+c_{26}+c_{36}+c_{35}
\end{equation}%
The dimers can be considered freely-jointed rigid rods, owing to the
flexibility of the DNA\ junctions which constitute the vertex. \ 

In general the free energy $\varepsilon _{ij}$ is equal to the hybridization
free energy $\Delta G_{ij}=\Delta H_{ij}-T\log (\Delta S_{ij})$ for joining
rods $R_{i}$ and $R_{j}$. \ These free energies will depend on the
particular choice of the DNA nucleotide sequences $\{A_{i},B_{j},C_{k},D_{l}%
\}$. \ In what follows $\varepsilon $ denotes the average hybridization free
energy $\left\langle \varepsilon _{ij}\right\rangle $. \ 

The reasoning for the trimer structures is largely the same. \ To write the
hybridization free energies for the $n$-mers compactly, we label them by the
indices for the rods which constitute the structure. \ For example, for the
trimer composed of rods $R_{1}$, $R_{2}$, and $R_{5}$ the effective free
energy is written $\varepsilon _{125}$. \ Looking at Fig. \ref{cagecoloring}%
, $\varepsilon _{125}$ can be decomposed into a sum of hybridization free
energies for joining two arms at a vertex, i.e. $\varepsilon
_{125}=\varepsilon _{15}+\varepsilon _{25}$. \ The same decomposition can be
done for all of the $n$-mers. \ We adopt the same notation for the
concentration of the structures. \ The concentration of the trimer $c_{125}$
formed by the reaction $R_{1}+R_{2}+R_{5}$ is%
\begin{equation}
c_{125}=\frac{c_{1}c_{2}c_{5}}{c_{o}^{2}}\exp \left( \frac{-\varepsilon
_{125}}{T}\right)
\end{equation}

For some of the DNA structures there is one additional complication. \ For
any diagram which contains a closed loop, we must calculate the change in
configurational entropy associated with making the connection which closes
the loop. \ In these cases we calculate the overlap density $c_{eff}$ which
relates the effective hybridization free energy $\widetilde{\varepsilon }$
to the bare hybridization free energy $\varepsilon $ (\cite{statmech},\cite%
{LicataPhD}). \ 
\begin{equation}
\exp \left( \frac{-\widetilde{\varepsilon }}{T}\right) =\frac{c_{eff}}{c_{o}}%
\exp \left( \frac{-\varepsilon }{T}\right)
\end{equation}%
\ 
\begin{equation}
c_{eff}=\frac{\int P(\mathbf{r}_{1},\mathbf{r})P(\mathbf{r}_{2},\mathbf{r}%
)d^{3}\mathbf{r}}{\left( \int P(\mathbf{r},\mathbf{r}^{\prime })d^{3}\mathbf{%
r}\right) ^{2}}
\end{equation}%
Here $P(\mathbf{r},\mathbf{r}^{\prime })$ is the probability distribution
for the chain of DNA linkers which starts at $\mathbf{r}^{\prime }$ and ends
at $\mathbf{r}$. \ The canonical example is the conversion of a trimer which
is a chain of three freely-jointed links into a closed triangle. \ For rigid
DNA linkers each of length $L$ the result is quite simple. \ 
\begin{equation}
c_{eff}=\frac{1}{8\pi L^{3}}
\end{equation}%
Details for the calculation are provided in an appendix. \ 

\begin{figure}[tbp]
\includegraphics[width=4.6138in,height=3.4714in]{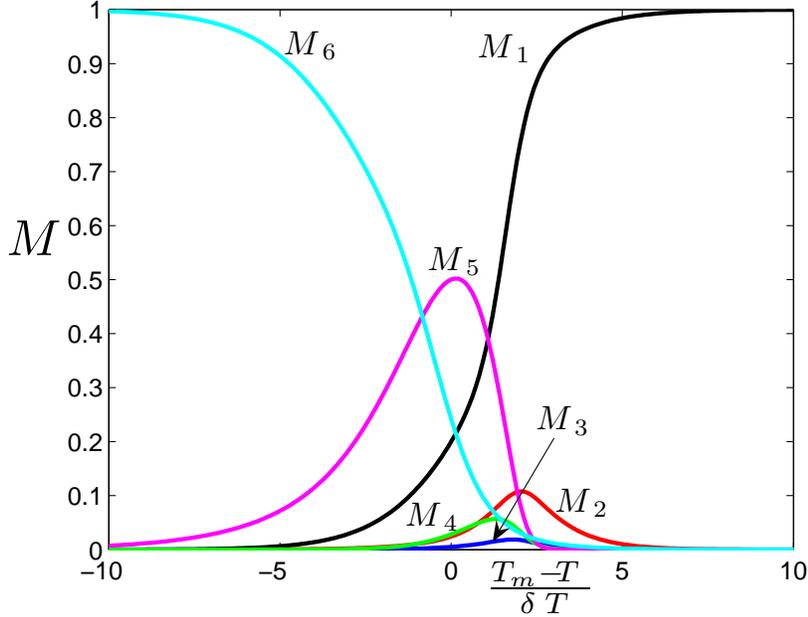}
\caption{(Color online).  The mass fraction for the
partially assembled cages which can form in the absence of the nanoparticle.
\ $M_{1}$, $M_{2}$, $M_{3}$, $M_{4}$, $M_{5}$, and $M_{6}$ are the mass
fractions for the monomers, dimers, trimers, tetramers, $5$-mers, and $6$%
-mer respectively. \ In the plot $c_{i}^{tot}=1nM$. \ }
\label{cagemeltnew}
\end{figure}

Continuing the enumeration procedure for the tetramers, $5$-mers, and $6$%
-mer, we can write down expressions for the concentration of all the DNA
structures which can form in the absence of the particle. \ Writing down the
equations for conservation of DNA linkers results in a system of $6$ coupled
polynomial equations of order $6$ in the concentrations of monomers $c_{j}$.
\ Here $c_{j}^{tot}$ is the total initial concentration of linkers of type $%
j $. \ 
\begin{eqnarray}
c_{1}^{tot}
&=&c_{1}+c_{15}+c_{16}+c_{125}+c_{146}+c_{126}+c_{135}+c_{136}+c_{156}+c_{1235}+c_{1236}
\\
&&+c_{1246}+c_{1346}+c_{1256}+c_{1356}+c_{1456}+c_{12346}+c_{12356}+c_{12456}+c_{13456}+c_{123456}
\notag \\
c_{2}^{tot}
&=&c_{2}+c_{25}+c_{26}+c_{125}+c_{236}+c_{126}+c_{235}+c_{246}+c_{256}+c_{1235}+c_{1236}+c_{1246}
\\
&&+c_{2346}+c_{1256}+c_{2356}+c_{2456}+c_{12346}+c_{12356}+c_{12456}+c_{23456}+c_{123456}
\notag \\
c_{3}^{tot}
&=&c_{3}+c_{35}+c_{36}+c_{236}+c_{135}+c_{136}+c_{235}+c_{346}+c_{356}+c_{1235}+c_{1236}+c_{1346}
\\
&&+c_{2346}+c_{2356}+c_{1356}+c_{3456}+c_{12346}+c_{12356}+c_{13456}+c_{23456}+c_{123456}
\notag \\
c_{4}^{tot}
&=&c_{4}+c_{46}+c_{146}+c_{246}+c_{346}+c_{1246}+c_{1346}+c_{2346}+c_{1456}+c_{2456}+c_{3456}
\\
&&+c_{12346}+c_{12456}+c_{13456}+c_{23456}+c_{123456}  \notag \\
c_{5}^{tot}
&=&c_{5}+c_{15}+c_{25}+c_{35}+c_{125}+c_{135}+c_{156}+c_{235}+c_{256}+c_{356}+c_{1235}+c_{1256}
\\
&&+c_{2356}+c_{1356}+c_{1456}+c_{2456}+c_{3456}+c_{12356}+c_{12456}+c_{13456}+c_{23456}+c_{123456}
\notag \\
c_{6}^{tot}
&=&c_{6}+c_{16}+c_{26}+c_{36}+c_{46}+c_{146}+c_{236}+c_{126}+c_{136}+c_{156}+c_{246}+c_{256}+c_{346}
\\
&&+c_{356}+c_{1236}+c_{1246}+c_{1346}+c_{2346}+c_{1256}+c_{2356}+c_{1356}+c_{1456}+c_{2456}
\notag \\
&&+c_{3456}+c_{12346}+c_{12356}+c_{12456}+c_{13456}+c_{23456}+c_{123456} 
\notag
\end{eqnarray}%
By solving these equations for the monomer concentrations we can plot the
melting profile (see Fig. \ref{cagemeltnew}). \ The plot is for the
symmetrical case $\varepsilon _{16}=\varepsilon _{46}=\varepsilon
_{15}=\varepsilon _{25}=\varepsilon _{26}=\varepsilon _{36}=\varepsilon
_{35}\equiv \varepsilon $. \ The results are plotted in terms of the
dimensionless variable $(T_{m}-T)/\delta T$ defined as:%
\begin{equation}
\frac{\left( T_{m}-T\right) }{\delta T}=\frac{\varepsilon }{T}-\log \left( 
\frac{\sum_{i}c_{i}^{tot}}{4c_{o}}\right)
\end{equation}%
Here $\delta T$\ is the width of the melting transition 
\begin{equation}
\delta T=\frac{T}{\Delta S+\log \left( \frac{\sum_{i}c_{i}^{tot}}{4c_{o}}%
\right) }\text{.}
\end{equation}%
$T_{m}$ is the melting temperature (neglecting the trimers and higher order
structures) for which the fraction of rods in the dimer phase $F=\left( 2%
\overline{C}_{2}\right) /(\overline{C}_{1}+2\overline{C}_{2})=1/2$. \ 
\begin{equation}
T_{m}=\frac{\Delta H}{\Delta S+\log \left( \frac{\sum_{i}c_{i}^{tot}}{4c_{o}}%
\right) }
\end{equation}%
The concentrations for of all the $n$-mers $\overline{C}_{n}$ are 
\begin{eqnarray}
\overline{C}_{1} &=&\sum_{i}c_{i} \\
\overline{C}_{2} &=&\sum_{j>i}c_{ij} \\
\overline{C}_{3} &=&\sum_{k>j>i}c_{ijk} \\
\overline{C}_{4} &=&\sum_{l>k>j>i}c_{ijkl} \\
\overline{C}_{5} &=&\sum_{m>l>k>j>i}c_{ijklm} \\
\overline{C}_{6} &=&\sum_{n>m>l>k>j>i}c_{ijklmn}
\end{eqnarray}%
where each index runs over the set $\{1,2,3,4,5,6\}$. \ For the summations
it is understood that the set of indices must form a connected diagram. \
For example, the term $c_{1234}$ does not appear in the sum for $\overline{C}%
_{4}$ since the vertex architecture (see Fig. \ref{cagecoloring}) stipulates
that this diagram represents two disconnected dimers, $c_{12}$ and $c_{34}$.
\ The mass fraction of the $n$-mers $M_{n}$ is then defined as 
\begin{equation}
M_{n}=\frac{n\overline{C}_{n}}{\sum\limits_{k=1}^{6}k\overline{C}_{k}}.
\end{equation}%

At low temperatures the dominant structure is the $6$-mer, which is the
fully assembled cage except for the binding of rod $R_{4}$ at vertex $D$ of
the cage. \ With this information at hand, we are now in a position to
determine the melting profile for the full system, DNA linkers together with
the DNA grafted nanoparticles. \ 

\section{DNA-caged particles}

In this section we determine the concentration of nanoparticles decorated
with DNA structures. \ We can determine the concentration of particles
decorated with DNA structures by applying the same rules for the chemical
potential as before. \ For example, consider decorating a free particle with
concentration $c_{p}$ with the monomer $c_{1}$. \ We have%
\begin{equation}
T\log \left( \frac{c_{p}}{c_{o}}\right) +T\log \left( \frac{c_{1}}{c_{o}}%
\right) =T\log \left( \frac{c_{p1}}{c_{o}}\right) +\widetilde{g}_{1}
\end{equation}%
The effective binding energy $\widetilde{g}_{1}$ has two contributions. \
The first comes from the hybridization free energy $g_{1}$ of the DNA arms
on the particle hybridizing with the ssDNA ends of rod $R_{1}$. \ As before
these hybridization free energies can be decomposed as a sum of
contributions from joining two arms at a vertex. \ Analogous to the
definition of $\varepsilon $, we let $g$ denote the average hybridization
free energy for joining two rods at the vertex, one of which came from the
DNA grafted on the nanoparticle surface. \ \ 

The second contribution is an entropic contribution associated with
localizing the DNA structure on the surface of the particle. \ Since there
are $N_{arms}$ DNA strands grafted onto the particle surface, there is a
combinatorial factor associated with the number of ways to make the first
connection between the particle and the DNA\ structure. \ Let $\sigma =$ $%
N_{arms}/(4\pi r^{2})$ be the average areal grafting density of DNA\ on the
nanoparticle surface for a particle of radius $r$, and $h\simeq 1nm$ a
localization length. \ The entropic contribution can be estimated in terms
of the concentration $\psi =\sigma /h$ which relates $\widetilde{g}$ to $g$
in the following manner. \ 
\begin{equation}
\widetilde{g}=(2\delta _{R_{1}}+2\delta _{R_{3}}+\delta _{R_{4}})g-T\log %
\left[ N_{arms}\left( \frac{\psi }{c_{o}}\right) ^{N-1}\right]
\end{equation}%
The factor $\delta _{R_{j}}=1$ if rod $R_{j}$ is present in the structure,
and $\delta _{R_{j}}=0$ otherwise. \ Here $N$ is the number of vertices of
the cage to which the nanoparticle is bound. \ In our example case we have $%
N=2$. \ Putting everything together we have 
\begin{eqnarray}
c_{p1} &=&\frac{c_{p}c_{1}}{c_{o}}\exp \left( \frac{-\widetilde{g}_{1}}{T}%
\right) \\
\widetilde{g}_{1} &=&2g-T\log \left( N_{arms}\frac{\psi }{c_{o}}\right)
\end{eqnarray}

\begin{figure}[tbp]
\includegraphics[width=4.6138in,height=3.4714in]{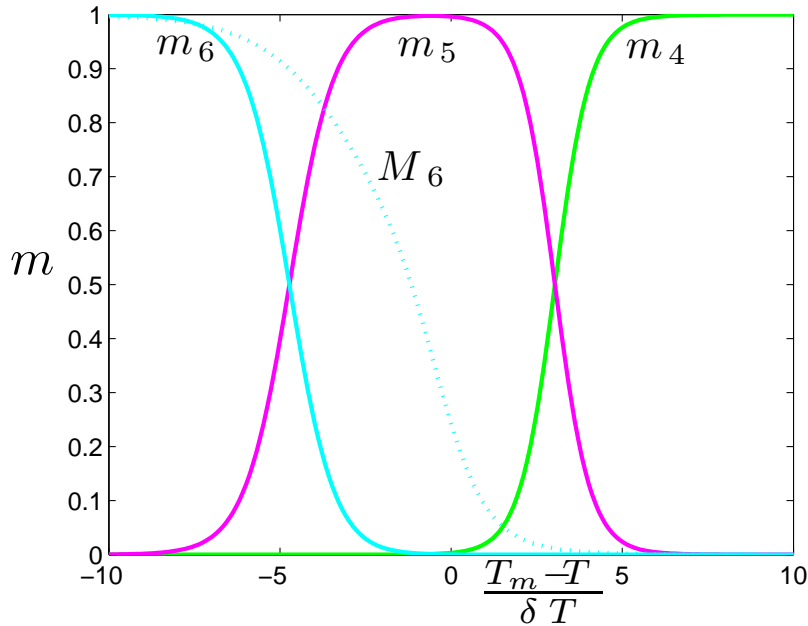}
\caption{(Color online). The mass fraction $m_{n}$ for nanoparticles decorated with $n$-mers. \
Note that the fully assembled tetrahedral cage surrounding the particle, $%
m_{6}$, is the dominant equilibrium structure for low temperatures. \ For
comparison the mass fraction of the cage in the absence of the particles $%
M_{6}$ is also plotted. \ In the plot $g=\protect\varepsilon $. \  }
\label{cagepartmeltnew}
\end{figure}

The same general procedure can be applied to decorating the particles with
all of the DNA structures considered in the previous section, making sure to
take into account the vertex architecture. \ For example, we cannot decorate
a particle with the dimer $c_{25}$ since at each of the vertices the DNA
arms which come from the particle cannot directly hybridize to the arms
which come from the dimer. \ The concentration of particles decorated with $%
n $-mers is $c_{p}^{(n)}$. \ 
\begin{eqnarray}
c_{p}^{(0)} &\equiv &c_{p} \\
c_{p}^{(1)} &=&c_{p}\sum\limits_{i}\frac{c_{i}}{c_{o}}\exp \left( \frac{-%
\widetilde{g}_{i}}{T}\right) \\
c_{p}^{(2)} &=&c_{p}\sum\limits_{j>i}\frac{c_{ij}}{c_{o}}\exp \left( \frac{-%
\widetilde{g}_{ij}}{T}\right) \\
c_{p}^{(3)} &=&c_{p}\sum\limits_{k>j>i}\frac{c_{ijk}}{c_{o}}\exp \left( 
\frac{-\widetilde{g}_{ijk}}{T}\right) \\
c_{p}^{(4)} &=&c_{p}\sum\limits_{l>k>j>i}\frac{c_{ijkl}}{c_{o}}\exp \left( 
\frac{-\widetilde{g}_{ijkl}}{T}\right) \\
c_{p}^{(5)} &=&c_{p}\sum\limits_{m>l>k>j>i}\frac{c_{ijklm}}{c_{o}}\exp
\left( \frac{-\widetilde{g}_{ijklm}}{T}\right) \\
c_{p}^{(6)} &=&c_{p}\sum\limits_{n>m>l>k>j>i}\frac{c_{ijklmn}}{c_{o}}\exp
\left( \frac{-\widetilde{g}_{ijklmn}}{T}\right)
\end{eqnarray}%
If $c_{p}^{tot}$ is the total initial particle concentration, we can write
the equation for particle conservation in the following form:%
\begin{equation}
c_{p}^{tot}=\sum_{n=0}^{6}c_{p}^{(n)}+O(c_{p}^{2})  \label{conservation}
\end{equation}%
This equation is then solved to determine the concentration of free
particles $c_{p}$ and hence the concentration for particles decorated with
DNA\ structures. \ The mass fraction $m_{n}$ for particles decorated with $n$%
-mers is%
\begin{equation}
m_{n}=\frac{c_{p}^{(n)}}{\sum\limits_{k=0}^{6}c_{p}^{(k)}}.
\end{equation}%
The results for the mass fraction are plotted (see Fig. \ref{cagepartmeltnew}%
) for the case $g=\varepsilon $. \ For low temperatures the dominant structure in solution is our
desired nanoblock, a DNA-caged particle. \

It is of crucial interest for the experimental feasibility of the proposal
that the tetrahedrally caged particle is the dominant structure close to
room temperature. \ We can see from Fig. \ref{cagepartmeltnew} that the
caged particle is the dominant equilibrium structure for $(T_{m}-T)/\delta
T\leq -5$. \ This in turn determines the standard enthalpy $\Delta H$ and
the standard entropy $\Delta S$ for the hybridization between two DNA arms
at the vertex. \ We find that $\Delta H\simeq -100$ kcal/mol and $\Delta
S\simeq -270$ cal K$^{-1}$ mol$^{-1}$. \ For DNA rods with concentration $%
c_{i}^{tot}=1nM$ in a $0.2M$ NaCl solution this gives $T_{m}\simeq 35^{o}C$
and $T\simeq 25^{o}C$. \ This information can be used to determine the
number of DNA bases in each arm of Fig. \ref{dnajunction} (i.e. the length
of the ssDNA ends on the rods). \ Using the average nearest neighbour
parameters of \cite{DNAmelting}, the DNA arms should be $28$ base pairs
long. \ Hence each of the dsDNA arms of the $4$ arm DNA junction is $14$
base pairs long. \ 

The essential result is that there is an experimentally accessible regime
for which the dominant equilibrium structure is the DNA-caged particle. \ In
this regime, $(T_{m}-T)/\delta T\leq -5$, we do not expect the assembly of
the DNA-caged particle to be kinetically limited. \ The reason is that in
this regime, even in the absence of the particle the dominant structure is
the $6$-mer, which is then trivially converted into the DNA-caged particle.
\ 

\section{DNA-particle parasites}

\begin{figure}[tbp]
\includegraphics[width=5.0548in,height=3.8009in]{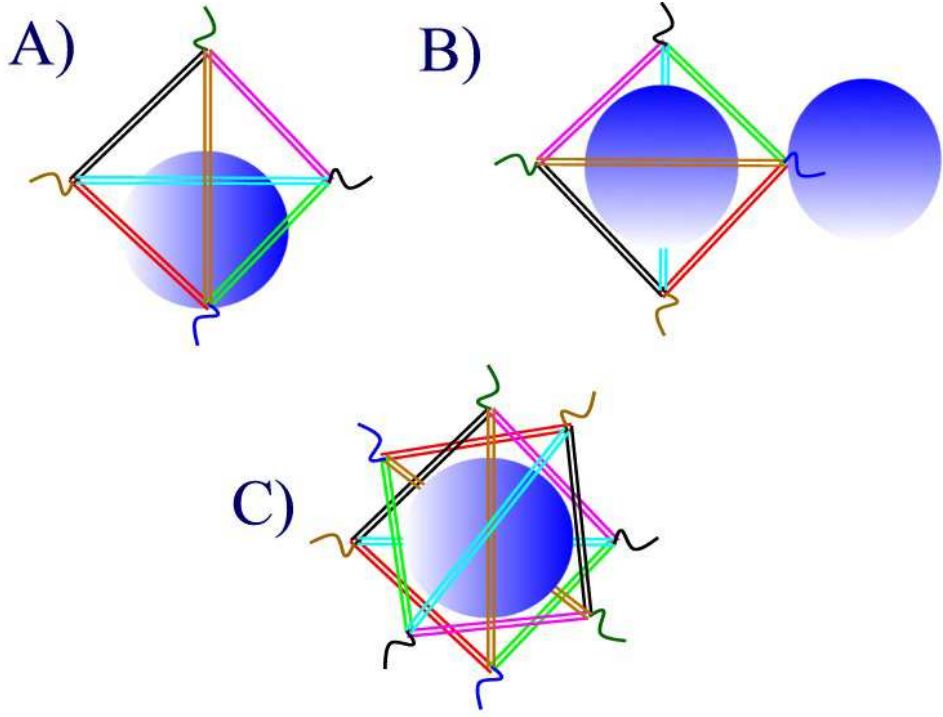}
\caption{(Color online). Some of the undesired structures which can form. \ A) One particle
attached to the outside of a cage. \ B) Two particles attached to the same
cage. \ C) One particle surrounded by two cages. \ }
\label{parasite}
\end{figure}

In this section we pause to discuss some other DNA-particle structures
(parasites) which could potentially decrease the overall yield of our
nanoblock (see Fig. \ref{parasite}). \ One competing structure is the
particle attached to the outside of a DNA cage. \ In this case the particle
can bind to at most three of the tetrahedron vertices. \ As a result, the
equilibrium yield of this structure will be proportional to the yield of the
DNA-caged particle, but suppressed by a factor of the Boltzmann weight for
the missing connection $\exp \left( \frac{-g}{T}\right) $, and
thus negligible for low temperature. \ 

We should also consider the stability of our nanoblock with respect to two
particle structures, i.e. including terms of order $c_{p}^{2}$ in Eq. \ref%
{conservation}. \ For example one can consider a particle pair attached to
the same DNA cage. \ Building such a structure does not necessarily cost
binding energy with respect to the DNA-caged particle. \ However, there is
still a cost is associated with the loss in translational entropy of the
free particle $T\log \left( \frac{c_{p}}{c_{o}}\right) $. \ As a result
these particle pairs will be relatively rare and should not decrease the
overall yield of the DNA-caged particle. \ 

In principle, a particle may have several DNA cages assembled around it. \
As it turns out, these are the most problematic of the competing structures.
\ If we denote the concentration of the cage $c_{123456}\equiv c_{cage}$ and
the free energy for the particle binding to the cage $\widetilde{g}_{123456}=%
\widetilde{g}_{cage}$ then concentration of particles with $m$ cages $%
C_{p}^{(m)}$ is 
\begin{equation}
C_{p}^{(m)}=c_{p}\left( \frac{c_{cage}}{c_{o}}\right) ^{m}\exp \left( \frac{%
-m\left[ \widetilde{g}_{cage}+(m-1)\alpha \right] }{T}\right)
\end{equation}%
Here $\alpha $ is an energetic parameter which characterizes the interaction
between two cages attached to the same particle. \ Since the dsDNA rods (the
cage edges) are charged, this interaction is presumably dominated by the
electrostatic repulsion of the rods. \ Within the Debye-Huckel approximation
this problem has been treated (\cite{rodinteraction},\cite{DNArod},\cite%
{zetapotential}). \ The electrostatic energy $E(R,\theta )$ of two rods
separated by a minimum center to center distance $R$ which make an angle $%
\theta $ when viewed along $R$ is 
\begin{equation}
\frac{E(R,\theta )}{T}=\left( \frac{2\pi \lambda _{B}}{\kappa l^{2}}\right) 
\frac{\exp (-\kappa R)}{\sin \theta }
\end{equation}%
Here the rods have the same effective linear charge density $\nu =\frac{e}{l}
$, $\lambda _{B}=\frac{e^{2}}{\epsilon T}$ is the Bjerrum length, and $%
\kappa ^{-1}=1/\sqrt{4\pi \lambda _{B}n}$ is the Debye screening length for
monovalent counterions of concentration $n$. \ Assuming that the
electrostatic energy for the cage-cage interactions can be expressed in
terms of a pairwise sum of contributions from rod-rod interactions we have 
\begin{equation}
\alpha \simeq \frac{N_{c}}{2}\left\langle E(R,\theta )\right\rangle .
\end{equation}%
Here the angular brackets denote the average, and $N_{c}$ is the number of
rod-rod contacts between two cages. \ An energetically favorable orientation
of the cages has $N_{c}=6$. \ To suppress the formation of particles with
two cages, we require the following ratio to be small 
\begin{equation}
\frac{C_{p}^{(2)}}{C_{p}^{(1)}}=\frac{c_{cage}}{c_{o}}\exp \left( \frac{-[%
\widetilde{g}_{cage}+\alpha ]}{T}\right) \ll 1  \label{suppress}
\end{equation}%
The electrostatic energy can be quite significant. \ For perpendicular
orientations of the dsDNA rods reference \cite{DNArod}\ reports a contact
potential $E\left( d,\frac{\pi }{2}\right) \simeq 50T$ (here $d\simeq 2.4nm$
is the dsDNA diameter) in $n=0.005M$ NaCl. \ At fixed temperature and salt
concentration, Eq. \ref{suppress} imposes a condition on the DNA\ linker
concentration which must be met in order to suppress the assembly of more
than one cage around the particle. \ If the DNA linker concentration is not
too high, and the salt concentration fairly low, the assembly of more than
one cage around the particle can be prevented. \ 

\section{Hierarchical self-assembly}

The DNA-caged nanoparticles in this paper are interesting nanoblocks in a
hierarchical self-assembly scheme. \ Part of their usefulness stems from the
fact that interactions between nanoblocks are highly anisotropic. \ Recall
that at each vertex of the DNA\ cage there is a unique ssDNA sequence
available for hybridization. \ As a result two nanoblocks can be made to
interact in a very specific manner by introducing another set of
vertex-vertex DNA\ linkers. \ Moreover, the number of these vertices is
explicitly determined by the cage architecture, which translates into a well
defined ``valence" for the interactions between nanoblocks. \

\begin{figure}[tbp]
\includegraphics[width=5.0548in,height=3.8009in]{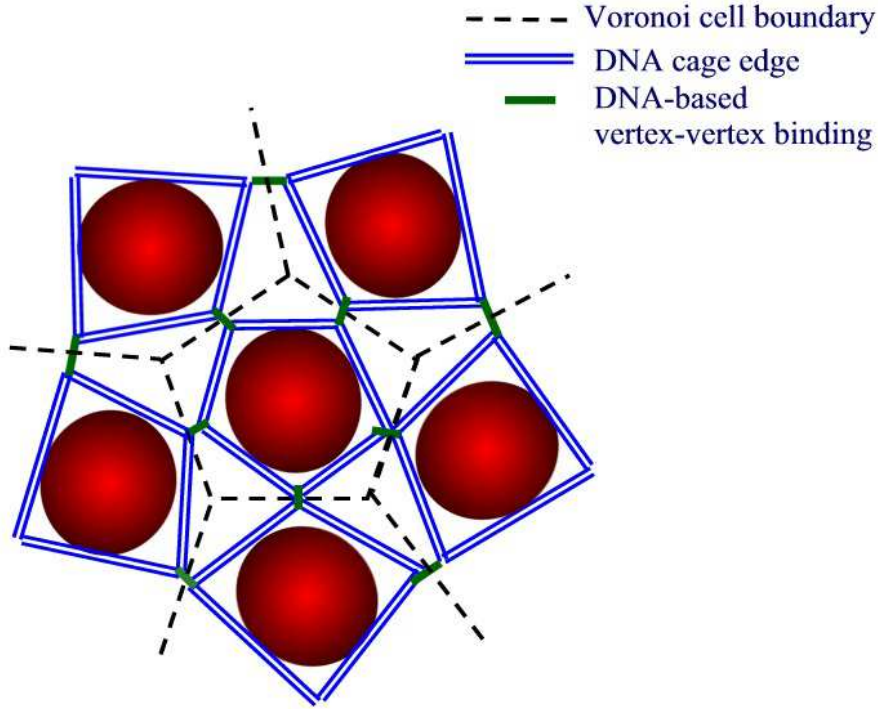}
\caption{(Color online). A simple example of the Voronoi scheme in two
dimensions. \ }
\label{Voronoi}
\end{figure}

Here we discuss a particular hierarchical self-assembly proposal where
DNA-caged particles are the natural building blocks. \ The basic proposal of
the Voronoi scheme is the following. \ Any target structure not necessarily
crystalline can be represented as a discrete set of points, i.e. the
location of particles in the structure. \ With this set, one performs the
Voronoi decomposition \cite{Voronoi}. \ The Voronoi cells for the structure
can be used to design the cages surrounding the particles in the following
manner. \ For any given particle, place a vertex at the midpoint between
that particle and each of its Voronoi neighbors. \ In this way we map the
target structure onto the set of caged particles, with certain pairs of
vertices to be connected. \ These vertices can be connected by introducing a
set of DNA\ linkers which perform the vertex-vertex binding. \ By
construction the target structure must be the ground state of the system. \
What remains to be seen is whether or not the target structure is favored
kinetically. \ A task for future research is a detailed consideration of
this question using Monte Carlo simulations. \ 

\section{Conclusions}

In this paper we discussed a proposal to self-assemble DNA-caged particles.
\ The basic components are several types of rod-like dsDNA linkers with
ssDNA\ ends, and nanoparticles grafted with ssDNA. \ By designing the ssDNA
sequences appropriately, the dsDNA rods self-assemble into a cage
surrounding the particle. \ The edges of the cage are dsDNA, and the
vertices are multi arm DNA junctions. \ A particular implementation of this
idea was discussed for the self-assembly of tetrahedrally caged particles. \
We calculated the equilibrium yield of the DNA-caged particles and discussed
their stability with respect to alternative structures. \ At low
temperature, the nanoparticle surrounded by one cage is the dominant
equilibrium structure. \ Although the calculations were performed for a
tetrahedral cage geometry, the ideas are generally applicable to many types
of polyhedral cages. \ Each vertex of the cage with degree $V$ can be
constructed from a $V+1$ arm DNA branched junction. \ Such junctions have
been constructed with up to $12$ arms, which leads open the possibility of
much more complicated cages (\cite{fivearm},\cite{8arm}). \ A natural next
step would be to consider a particle inside a DNA cube, since the same
vertex architecture proposed in this paper would apply. \ 

We concluded by discussing the usefulness of DNA-caged particles in a
hierarchical self-assembly proposal. \ The Voronoi scheme maps the problem
of self-assembling a particular target structure onto a set of caged
particles for which the target structure is the ground state. \ The
experimental realization of self-assembled DNA-caged particles would
represent an important step towards realizing the technological potential of
DNA. \ 

Thanks to A. Klopper for help with the figures. \ This work was supported by the ACS Petroleum Research Fund (Grant PRF No.
44181-AC10). \ 
\bibliographystyle{achemso}
\bibliography{acompat,dna}

\section{Appendix: Overlap density $c_{eff}$}

In this appendix we calculate the overlap density $c_{eff}$ used to
determine the effective hybridization free energy for the DNA structures in
Fig. \ref{mers}. \ It is helpful to consider the associated problem of
determining the probability distribution for the end vector of a
freely-jointed chain made up of $N$ linkers, each of length $L$. \ The
probability distribution for the chain composed of one linker is simply. \ 
\begin{equation}
\rho _{1}(\mathbf{R})=\frac{\delta (R-L)}{4\pi L^{2}}
\end{equation}%
Therefore for a chain composed of $N$ such linkers we have:%
\begin{equation}
\rho _{N}(\mathbf{R})=\prod\limits_{j=1}^{N}\int d^{3}\mathbf{r}_{j}\rho
_{1}(\mathbf{r}_{j})\delta ^{3}\left( \sum\limits_{i=1}^{N}\mathbf{r}_{i}-%
\mathbf{R}\right)
\end{equation}%
The inverse Fourier transform of the probability distribution has a
particularly simple form \cite{statprop}. \ 
\begin{equation}
\widetilde{\rho }_{N}(\mathbf{k})=\int d^{3}\mathbf{R}\exp (i\mathbf{k\cdot R%
})\rho _{N}(\mathbf{R})=\left( \frac{\sin (kL)}{kL}\right) ^{N}
\end{equation}%
\begin{equation}
\rho _{N}(\mathbf{R})=\left( 2\pi \right) ^{-3}\int d^{3}\mathbf{k}\exp (-i%
\mathbf{k\cdot R})\widetilde{\rho }_{N}(\mathbf{k})
\end{equation}%
Working in spherical coordinates, performing the angular integration yields%
\begin{eqnarray}
\rho _{N}(\mathbf{R}) &=&\left( 2\pi \right) ^{-3}\int_{0}^{\infty
}k^{2}\left( \frac{\sin (kL)}{kL}\right) ^{N}dk\int d\Omega \exp (-ikR\cos
\theta ) \\
&=&\left( 2\pi \right) ^{-2}\frac{2}{R}\int_{0}^{\infty }k\sin (kR)\left( 
\frac{\sin (kL)}{kL}\right) ^{N}dk  \notag \\
&=&\frac{1}{2\pi ^{2}L^{3}}\int_{0}^{\infty }u^{2}j_{0}\left( \frac{Ru}{L}%
\right) \left[ j_{0}(u)\right] ^{N}du  \notag
\end{eqnarray}%
Here $j_{0}(z)=\frac{\sin z}{z}$ is the spherical Bessel function \cite%
{Bessel} of order $0$. \ 

We are now in a position to determine the overlap density $c_{eff}$ for the
triangle structure. \ Of particular interest for this calculation is (\cite%
{stochastic},\cite{randomflight}):%
\begin{equation}
\rho _{2}(\mathbf{R})=\frac{\theta (2L-R)}{8\pi L^{2}R}
\end{equation}%
Here $\theta (x)$ is the Heaviside step function. \ The overlap density for
the triangle is then calculated as%
\begin{eqnarray}
c_{eff} &=&\int d^{3}\mathbf{R}\int d^{3}\mathbf{r}\rho _{2}(\mathbf{R})\rho
_{1}(\mathbf{r})\delta ^{3}(\mathbf{R-r}) \\
&=&\frac{1}{8\pi L^{2}}\frac{1}{4\pi L^{2}}\int d^{3}\mathbf{R}\int d^{3}%
\mathbf{r}\frac{\theta (2L-R)}{R}\delta (r-L)\delta ^{3}(\mathbf{R-r}) 
\notag \\
&=&\frac{1}{32\pi ^{2}L^{4}}\int d^{3}\mathbf{r}\frac{\theta (2L-r)}{r}%
\delta (r-L)  \notag \\
&=&\frac{1}{8\pi L^{3}}  \notag
\end{eqnarray}

There are two remaining overlap densities which need to be calculated. \ One
of them is for forming an equilateral parallelogram, i.e. a diamond. \ In
Fig. \ref{mers} this structure is the third structure in the tetramer row. \
Forming this structure can be viewed as joining the ends of two chains, each
of which is composed of two links. \ 
\begin{eqnarray}
c_{eff} &=&\int d^{3}\mathbf{R}\int d^{3}\mathbf{r}\rho _{2}(\mathbf{R})\rho
_{2}(\mathbf{r})\delta ^{3}(\mathbf{R-r}) \\
&=&\left( \frac{1}{8\pi L^{2}}\right) ^{2}\int d^{3}\mathbf{R}\int d^{3}%
\mathbf{r}\frac{\theta (2L-R)}{R}\frac{\theta (2L-r)}{r}\delta ^{3}(\mathbf{%
R-r})  \notag \\
&=&\frac{1}{64\pi ^{2}L^{4}}\int d^{3}\mathbf{r}\frac{\theta (2L-r)}{r^{2}} 
\notag \\
&=&\frac{1}{8\pi L^{3}}  \notag
\end{eqnarray}%
Alternatively, one can calculate the overlap density for the diamond
structure as%
\begin{equation}
c_{eff}=\int d^{3}\mathbf{R}\int d^{3}\mathbf{r}\rho _{3}(\mathbf{R})\rho
_{1}(\mathbf{r})\delta ^{3}(\mathbf{R-r})=\frac{1}{8\pi L^{3}}
\end{equation}%
This is simply viewing the diamond as joining a chain of three links with a
chain of one link. \ The results are the same, as they must be. \ 

The remaining overlap density to be calculated is associated with making the
last connection in the tetrahedron. \ Assume that all the connections have
been made, except for the connection between rod $R_{4}$ and vertex $A$. \
Taking vertex $D$ as the origin, the position of vertex $A$ is the end
vector of a two segment chain, with each segment having length $l=\frac{L%
\sqrt{3}}{2}$. \ With one end of rod $R_{4}$ fixed at vertex $D$, the other
end must connect to vertex $A$. \ 
\begin{eqnarray}
c_{eff} &=&\int d^{3}\mathbf{R}\int d^{3}\mathbf{r}\rho _{2}(\mathbf{R})\rho
_{1}(\mathbf{r})\delta ^{3}(\mathbf{R-r}) \\
&=&\frac{1}{8\pi l^{2}}\frac{1}{4\pi L^{2}}\int d^{3}\mathbf{R}\int d^{3}%
\mathbf{r}\frac{\theta (2l-R)}{R}\delta (r-L)\delta ^{3}(\mathbf{R-r}) 
\notag \\
&=&\frac{1}{32\pi ^{2}l^{2}L^{2}}\int d^{3}\mathbf{r}\frac{\theta (2l-r)}{r}%
\delta (r-L)  \notag \\
&=&\frac{1}{8\pi l^{2}L}  \notag \\
&=&\frac{1}{6\pi L^{3}}  \notag
\end{eqnarray}

\end{document}